\documentclass[trackchanges]{aastex62}

\usepackage{CJK}
\usepackage{amsmath}
\usepackage{tikz}
\usetikzlibrary{arrows}
 \usepackage{booktabs}
 \usepackage{multirow}
\usepackage{tabularx}
\usepackage{booktabs,multirow}
\shorttitle{Late Phase}
\shortauthors{Zhou et al.}
\listofchanges
\begin{document}
\begin{CJK*}{UTF8}{gbsn}

\title{Extreme-ultraviolet Late Phase Caused by Magnetic Reconnection over Quadrupolar Magnetic Configuration in a Solar Flare}

\correspondingauthor{Zhenjun Zhou}
\email{zhouzhj7@mail.sysu.edu.cn}
\author[0000-0001-7276-3208]{Zhenjun Zhou(周振军)}
\affil{School of Atmospheric Sciences, Sun Yat-sen University, Zhuhai, Guangdong, 519000, China}
\author[0000-0003-2837-7136]{Xin Cheng}
\affil{School of Astronomy and Space Science, Nanjing University, Nanjing 210093, People's Republic of China}
\affil{Key Laboratory for Modern Astronomy and Astrophysics (Nanjing University), Ministry of Education, Nanjing 210093, People's Republic of China}
\author[0000-0001-6804-848X]{Lijuan Liu}	 
\affil{School of Atmospheric Sciences, Sun Yat-sen University, Zhuhai, Guangdong, 519000, China}
\author[0000-0001-9856-2770]{Yu Dai}
\affil{School of Astronomy and Space Science, Nanjing University, Nanjing 210093, People's Republic of China}
\affil{Key Laboratory for Modern Astronomy and Astrophysics (Nanjing University), Ministry of Education, Nanjing 210093, People's Republic of China}
\author[0000-0002-8887-3919]{Yuming Wang}
\affil{CAS Key Laboratory of Geospace Environment,Department of Geophysics and Planetary Sciences, University of Science and Technology of China, Hefei, Anhui 230026, China}
\affil{Synergetic Innovation Center of Quantum Information \& Quantum Physics, University of Science and Technology of China, Hefei, Anhui 230026, China}
\author[0000-0002-4721-8184]{Jun Cui}
\affil{School of Atmospheric Sciences, Sun Yat-sen University, Zhuhai, Guangdong, 519000, China}

\begin{abstract}
A second emission enhancement in warm coronal extreme-ultraviolet (EUV) lines (about 2-7 MK) 
during some solar flares is known as the EUV late phase. Imaging observations confirm that the late phase emission
originates from a set of longer or higher loops than the main flare loops. 
Nevertheless, some questions remain controversial: What is the relationship between these two loop systems? 
What is the heating source of late phase emission,
a heating accompany the main phase heating or occuring quite later? 
In this paper, we present clear evidence for heating source in a late-phase solar flare: magnetic reconnection of overlying field in a quadrupolar magnetic configuration.
The event is triggered by an erupted core structure that eventually leads to a coronal mass ejection (CME).
Cusp feature and its shrinkage motion high in the late-phase emission region are the manifestation of the later phase reconnection following the main flare reconnection.
Using the enthalpy-based thermal evolution of loops (EBTEL) model, we reasonably reproduce the late-phase emissions in some EUV lines. We suggest that a continuous additional heating is responsible for  the appearance of the elongated  EUV late phase.

\end{abstract}

\keywords{Sun: corona --- Sun: flares --- Sun: UV radiation}

\section{Introduction} \label{sec:intro}

A solar flare,  localised brightening in the solar atmosphere,  rapidly releases radiation over a wide range of electromagnetic spectrum. 
Beside the radiation enhancement,  sometimes the flare also accelerates high energy particles and  is accompanied with  magnetic flux ejected into the interplanetary space, which is manifested as a coronal mass ejection (CME). 

Solar flares are generally believed to be a result of magnetic reconnection - the merging of 
antiparallel magnetic fields and the consequent release of magnetic energy.
The free magnetic energy is convert into the kinetic energy of the CME and significant particle acceleration and plasma heating. 
When the impulsive flare heating is terminated, the plasma will cool down mainly by conduction and radiation.
Initially, the energy is transported downward mainly in the form of conductive flux or non-thermal particles and deposited in the transition region (TR) and chromosphere.
Due to the Coulomb collisions, heating, and strong ablation,
the ambient chromospheric plasma 
is strongly heated and expands upward, filling the coronal loops, known as the chromospheric evaporation\citep{Antiochos_Sturrock_1978}. 
As the loop temperature decreases and the density increases,
radiation gradually becomes dominant. As the plasma cools, pressure gradients drop to subhydrostatic values, and material drains from the corona \citep{Klimchuk_etal_2008}.

Observations show that flares typically experience three  stages: (1) a precursor phase, presenting a small emission enhancement at radio, H$\alpha$, UV to EUV wavelengths \citep[e.g.,][]{Bumba_Krivsky_1959, Martin_1980, Van_Hurford_1984, Cheng_etal_1985, Warren_Warshall_2001, Farnik_etal_2003, Contarino_etal_2003}; (2) an impulsive phase indicated by rapidly enhancement of a microwave burst and/or a hard X-ray burst that lasts from 100 to 1000 seconds in which protons and electrons are accelerated to energies over 1 MeV , and (3) a decay phase that lasts for several minutes or hours with the gradual increase and decrease of soft X-ray and EUV emissions\citep[and reference therein]{Hudson_2011}. 

However, recent observations revealed that in some solar flares, 
warm coronal emission lines (about $2-7$ MK) exhibit a second emission peak \citep{woods_etal_2011} as seen by the EUV 
Variability Experiment \citep[EVE;][]{Woods_etal_2012} on board the
Solar Dynamics Observatory \citep[SDO;][]{Pesnell_etal_2012}. 
This secondary peak, referred to the late phase and not subject to any stages mentioned above, lags behind the GOES X-ray peak for tens of minutes to hours.
\citet{woods_etal_2011} proposed four observational criteria to define an EUV late phase:
(1) a second peak in the warm emissions
(e.g., Fe XV and Fe XVI) after the GOES soft X-ray peak; 
(2) no significant enhancements in hot emissions (e.g., Fe XX) during
the second peak; 
(3) association with an eruptive event seen in
imaging observations; and 
(4) the existence of a second set of higher and
longer loops relative to the main flaring loops.
In general, the emission of an EUV late-phase flare is from the synthesis effect of thermal evolution from these two sets of loop systems.
The difference in cooling rate between different loops could discriminate the two peaks in warm coronal emission lines \citep{Liuk_etal_2013,Wang_etal_2016}. 
Yet, cooling process of another longer loop alone cannot be sufficient for some long delay (several hours) cases \citep{Hock_etal_2012}.

Nevertheless, how is the late phase loop system heated up  is still uncovered. Studies have explored the possibility of
additional heating \citep{woods_etal_2011,Hock_etal_2012,Dai_etal_2013,Sun_etal_2013}, as well as the long lasting
 cooling timescale \citep{Liuk_etal_2013,Sun_etal_2013}.
Using the enthalpy-based thermal evolution of
loops (EBTEL) model\citep{Klimchuk_etal_2008, Cargill_etal_2012}, \citet{Dai_etal_2018a} 
numerically simulated EUV late-phase flares under two main
mechanisms, i.e., long duration plasma cooling and additional heating mechanisms.  They found that the late phase peak occurs during the radiative cooling phase of the late-phase loop
if it is a long-lasting cooling process, while the additional heating  probably makes the late-phase peak take place in the conductive cooling phase.  They proposed that the shape of late phase light curves can be used to differentiate the two mechanisms.

Meanwhile,  statistical analysis \citep{woods_etal_2011} showed that the flares with late phase exhibit clustering phenomena in certain active region (AR),
implying a specific magnetic configuration of the ARs in which EUV late phase flares are preferentially produced. 
From direct EUV imaging observation, 
the basic 2D configuration of the magnetic topology is considered as a quadrupolar
configuration - typical symmetric   \citep{Hock_etal_2012} or asymmetric \citep{Liuk_etal_2013} quadrupolar
configuration. In reality, ARs often show a more complicated configuration. 
\citet{Sun_etal_2013} considered a fan-spine topology with a closed-field configuration, in which the late phase peak comes from
the cooling of large post-reconnection loops beside and above the compact fan. This magnetic field configuration has been further verified through 
observation \citep{Dai_etal_2013, Li_etal_2014, Masson_etal_2017}. 
The key ingredients pertinent to AR magnetic configuration with late phase can be simply summarized as a quadrupolar
configuration. This kind of topology distinguishes two sets of loop systems - the external late phase loop system and inner main phase loop system. \citet{Sun_etal_2013} suggested that the two systems and their evolutional trends are closely linked by the fan-spine topology in their case. While \citet{Hock_etal_2012} considered that these two loop systems have no direct linkage, and it is the CME that stretches the late phase loops to reconnect. In both of cases, during the late-phase stage, cusp-shaped structures are found above main flare region,  which is characteristic for ongoing
reconnection behind ejecta.  Particularly, when considering  additional heating, the quantitative results are more consistent with observations \citep{Sun_etal_2013}. 
 
Based on the above, the mechanism causing the late phase is still under debate.
In this paper, we study a limb flare event with a well-defined EUV late phase.  In section ~\ref{sec:Ins}, we introduce the instruments. In section ~\ref{sec:Obs}, we present observations 
and analysis of the event, which are followed by an EBTEL modeling of the late-phase emission in section~\ref{sec:mod}. Finally we  present summary and
conclusion in  section~\ref{sec:dis}.
 
\section{Instruments} \label{sec:Ins}
The data presented in this work primarily come from  the Atmospheric Imaging Assembly \citep[AIA;][]{Lemen_etal_2012},
which images the coronal plasma through 6 different EUV passbands. 
Among them, the 13.1 nm (Fe XXI, $\sim$ 10 MK),  9.4 nm (Fe XVIII, $\sim$ 6.4 MK),
and 33.5 nm passbands (Fe XVI, $\sim$ 2.5 MK)  are sensitive to hot(warm) plasma and can be used for thermal evolution analysis, and the other three passbands,  
21.1 nm (Fe XIV, $\sim$ 2.0 MK), 19.3 nm (Fe XII, $\sim$ 1.6 MK), 
and 17.1 nm (Fe IX, $\sim$ 0.7 MK) are largely sensitive to plasma 
cooler than 2.0 MK, e.g., bulk coronal plasma. 
The cadence of these EUV images is 12 s and the pixel size 
is $0 \farcs 6$ . 
Note that the AIA 13.1 nm passband also includes Fe VIII lines that are 
expected to be brighten up at relatively low coronal temperatures ($\sim$ 0.4 MK), 
and the AIA 19.3 nm passband contains a hot 
spectral line  Fe XXIV ($\sim$ 20 MK)\citep{Lemen_etal_2012}. 
The twin spacecraft of the Solar Terrestrial Relationships Observatory \citep[STEREO;][]{Kaiser_etal_2008} 
provide simultaneous 
multi-viewpoint from other two places. During the event, STEREO Ahead and Behind are about 
157 $^{\circ}$ west and 165 $^{\circ}$ east from the Sun-Earth
line along the ecliptic orbit around the Sun, while the SDO 
is situated near the Earth. The Extreme Ultraviolet Imager (EUVI) 
 on the two STEREO spacecraft observe the chromosphere and low 
corona in four EUV passbands, which enables us to reconstruct 
the 3D shape of coronal loops using triangulation technique. 
Here we choose the 19.5 nm (Fe XII, $\sim$ 1.5 MK) 
passbands from EUVI\_A,B and 19.3 nm (Fe XII, $\sim$ 1.6 MK) 
passband from SDO/AIA to display the evolution of the coronal 
loop in different angles and reconstruct their geometrical configuration. 
We use wavelet-enhanced images  of STEREO EUVI provided by Johns Hopkins University Applied Physics Laboratory(JHUAPL). This image-processing technique (wavelet-enhanced) gives better visual clarity than the standard EUVI images\citep{Stenborg_etal_2008}.

Moreover, we make use of EVE observations, which obtain integrated emission from the Sun over a wide wavelength range (0.1 nm to 105 nm)  with unprecedented 
spectral resolution of 0.1 nm, high temporal cadence (10s), and accuracy
of 20\% \citep{Woods_etal_2012}. 
EVE provides two set of level 2 data products including the ``line'' (EVL) 
product and the ``spectra'' (EVS) product, which are publicly available at 
 \url{http://lasp.colorado.edu/eve/data_access/evewebdataproducts/level2/}. 

\section{Observations and Results} \label{sec:Obs}
\subsection{Overview of the Event}
On 2014 April 25, a solar flare started at 00:17 UT and lasted over eight hours according 
to the GOES soft X-ray (SXR) 
flux (Figure~\ref{fig:figure1}(a)). The SXR flux( 1 -- 8 \AA) reached a maximum of 
$1.3 \times 10^{-4}$  Wm$^{-2}$, revealing an X1.3 class flare. At the time of the flare, the associated AR NOAA 12035  was located 
behind the west limb of the sun. 
The accompanied CME had an initial speed around 600 km s$^{-1}$ at a height
of $\sim$ 3R$_s$. However the velocity decreased to 300 km s$^{-1}$  when the CME reached 
20 R$_s$ according to the  \dataset[CME catalog]{cdaw.gsfc.nasa.gov/CME_list} 
based on the Large Angle Spectroscopic
 Coronagraph C2 and C3  observation \citep[LASCO;][]{Brueckner_etal_1995}. 
Despite at that time the magnetic field 
of the AR can not be measured, the magnetograms recorded 
three days prior to the eruption revealed a quadrupolar magnetic configuration
 on the photosphere. The onset of the event was believed to be caused by breakout reconnection\citep{Chen_etal_2016}. 
\subsection{Extremely Large EUV Late Phase} \label{subsec:Lat}
\begin{figure}[ht!]
\plotone{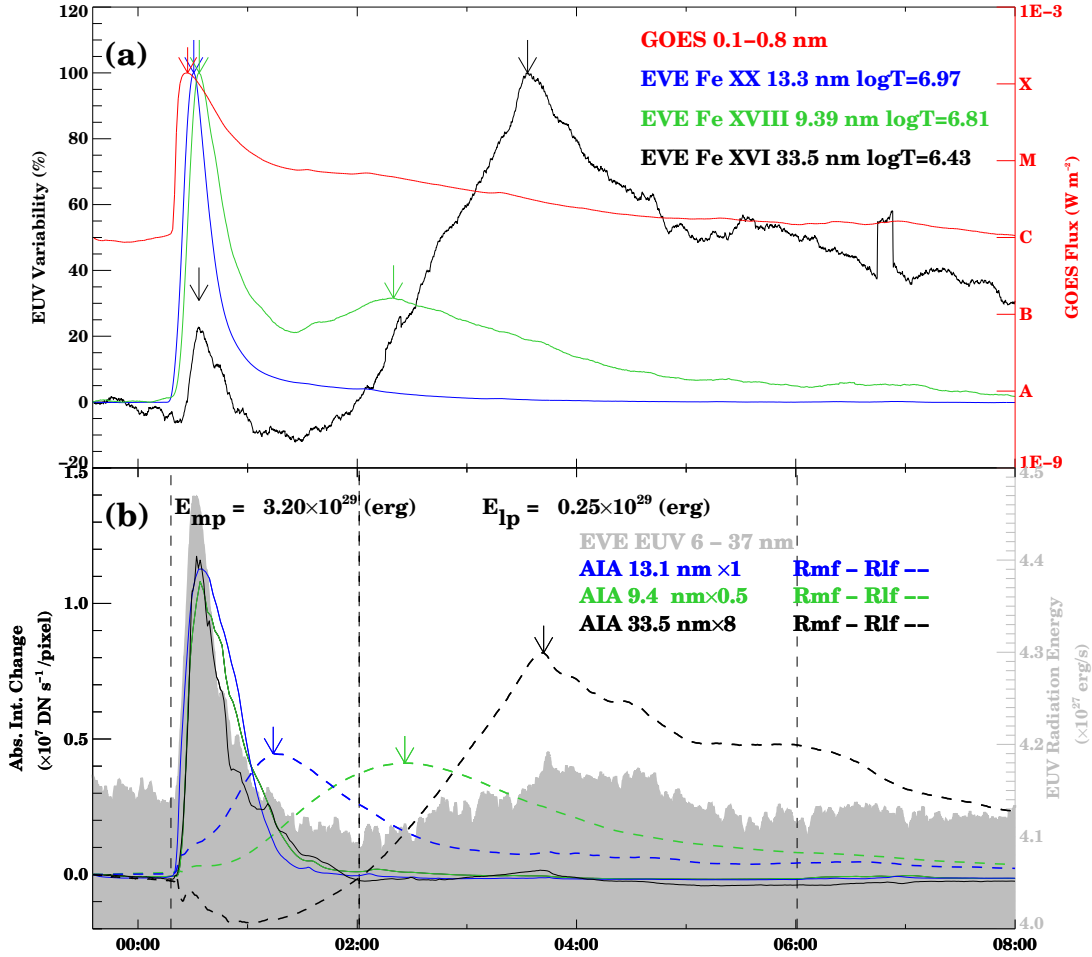}
\caption{Time profiles of each channel for the 2014 April 25 X1.3 flare. Panel (a) shows the background-subtracted irradiance in 3 EVE lines (normalized) and GOES 0.1 -- 0.8 nm flux. Panel (b) gives light curves from regions indicated by the boxes in figure~\ref{fig:figure2}, with the total EUV radiative energy-loss rate overplotted. The dashed lines come from the large box region in figure~\ref{fig:figure2}(i) and solid lines from small box. The color-coded arrows denote the peak time for the corresponding emission.\label{fig:figure1}}
\end{figure}

Three emission lines from the EVL data are selected to study the EUV evolution of this event with  the wavelength centered at  13.3 nm (Fe XX, $\sim$ 10 MK), 94 nm (Fe XVIII, $\sim$ 6.5 MK), and 33.5 nm (Fe XVI, $\sim$ 2.5 MK).  
These three EVE lines have complementary imaging observations from AIA. 
Subtracting the emission several minutes before the eruption, 
the temporal profiles of irradiance variability (normalized) in these lines
during the flare period are plotted as blue, green, and black
curves in Figure~\ref{fig:figure1}(a), respectively. The intensities of three EVE lines almost simultaneously enhance with the GOES SXR
flux. The delay of the first peak
(as pointed by the colored arrow) indicates the
cooling process of the flare's main phase.   Unlike the only one peak in GOES SXR and 13.3 nm profile,
a second peak appears in 9.4 nm and 33.5 nm. The intervals between the two peaks in 9.4 nm
and 33.5 nm are 106 and 180 minutes, respectively.  These secondary emission peaks in warm EUV lines (about 2-7 MK) 
 are signatures of the EUV late phase. Furthermore, in the warm 33.5 nm line profile,  the second peak 
is much stronger than the first peak, indicative of an extremely large EUV late phase \citep{Liu_etal_2015}. The 
flux ratio of the second peak to the first peak is around 5, much higher than the average value of 0.8 in statistics \citep{woods_etal_2011}.

\begin{table}[!htbp] 
\centering 
 \caption{Properties of the EUV Light Curves of the 2014 April 25 X1.3 Flare{$^\dag$}}\label{tb:table1}

\begin{tabular}{@{\extracolsep{1pt}} cccccccc} 

\toprule
{\centering Line/Passband}
& Ion & Temperature
  & \multicolumn{2}{c}{Peak Time (UT)} &&
\multicolumn{2}{c}{Delay Time (minutes)} \\ 
\cmidrule{4-5} \cmidrule{7-8}
(nm)& &(MK)& {\centering Main Phase} & {Late Phase} && {Main Phase} & {Late Phase}  \\
\midrule

SXR (0.1-0.8)    & *           & $>$10     & 00:27:10    & *        &   & 10 & *    \\ 
MEGSA (6-37)     & *           & *         & 00:30:32    & 03:44:32 &   & 13 & 207  \\ 
13.3             & Fe XX       & 10        & 00:30:32    & *        &   & 13 & *    \\ 
9.4              & Fe XVIII    &6.5        & 00:33:32    & 02:19:52 &   & 16 & 122  \\ 
33.5             & Fe XVI      & 2.5       & 00:33:22    & 03:33:12 &   & 16 & 196  \\ 
$13.1^a$         & Fe XXI      & 10        & 00:35:00    & 01:14:00 &   & 18 & 57   \\ 
$9.4^a$          & Fe XVIII    & 6.4       & 00:34:00    & 02:26:02 &   & 17 & 129  \\ 
$33.5^a$         & Fe XVI      & 2.5       & 00:32:00    & 03:42:00 &   & 15 & 205  \\ 

\hline \\[-1.8ex] 
\end{tabular}\\
\raggedright
$^\dag$  In this table, the delay time is calculated from peak time to start time (00:17 UT) of the eruption. Superscript ``a'' means the light curve from integrated average over the area of the respective box in AIA image. AIA 13.1 nm passband also contains a cold line (Fe VIII, 0.4 MK), in this study, the enhanced emission should be from the hot plasma during its peak time. 

\end{table}

The EVS spectral data from MEGS-A instrument provide continuous measurement for spectrum
irradiance variability of the flare within the wavelength range
of 6-37 nm. Here, the routine eve\_integrate\_line.pro from SolarSoftware (SSW) package is employed to integrate irradiance over this wavelength range using midpoint rule:
\begin{equation}
E(t)=K\sum_{\lambda_{min}}^{\lambda_{max}} \Delta\lambda I_\lambda(t)\,,
\label{eqn:eqn1}
\end{equation}
where $\Delta\lambda$ is the wavelength bin, and  $K$ is a conversion factor (=$10^7 \times 2 \times \pi \times$ (1AU)$^2$) that transforms the irradiance flux in units of J m$^{-2}$ at 1 AU to the radiative loss rate in units of erg s$^{-1}$ at the Sun assuming a uniform
angular distribution of flare energy release \citep{Woods_etal_2006}.  
The radiative energy-loss rate in this EUV range is calculated according to Equation~\eqref{eqn:eqn1}\citep{Liu_etal_2015} and
displayed with shading area in figure~\ref{fig:figure1}(b). It is interesting to note 
that there are also two peaks in the profile of the EUV radiative energy-loss rate. 
The vertical dashed lines bound the impulsive and the gradual phases from the two enhancements of Fe xvi 33.5 nm line irradiance, 
the former is from the onset of the impulsive phase at 00:17 UT to its conclusion at 02:00 UT, the latter includes the late-phase EUV enhancement between 02:00 UT and 06:00 UT.
We also calculate the total flare EUV output in these two periods
by simply integrating the emission with the background emission subtracted. 
In spite of a higher late-phase peak in the 33.5 nm line,
the EUV radiation output releases much less energy ($0.25 \times 10^{29}$ erg) during the flare late phase than that ($3.2 \times 10^{29}$ erg) during the flare main phase.

\begin{figure}[ht!]
\plotone{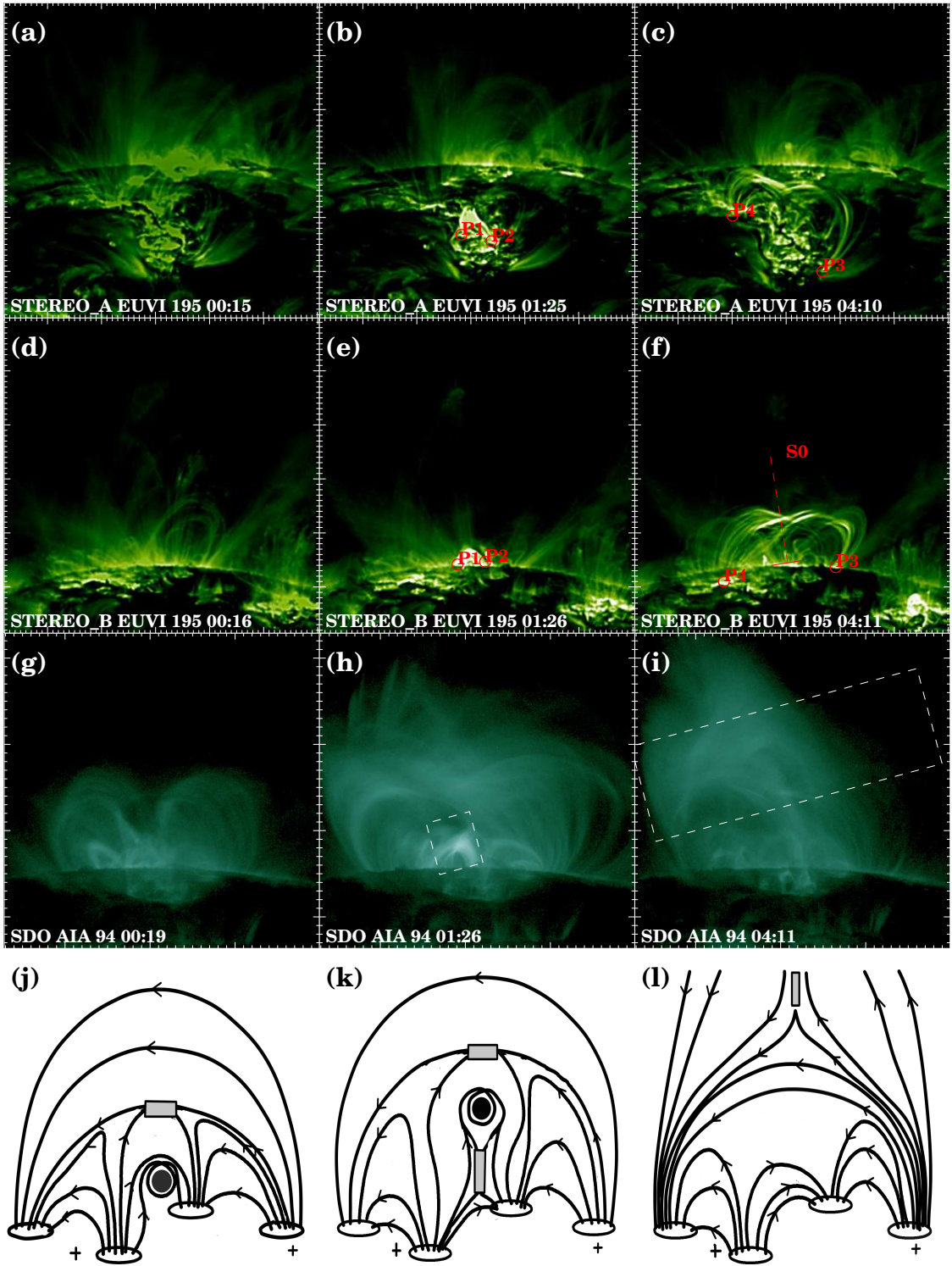}
\caption{Three stages of this solar flare from different view angles. 
The first three rows are the observations from EUVI\_A 19.5 nm,  EUVI\_B 19.5 nm and AIA 9.4 nm, respectively.  The last row shows schematic interpretation. All the observations are rotated with the eruption direction pointing to the top. The first column ((a),(d),(g) and (j)) shows the initial phase of the solar flare.  The middle column  dispays the main phase and last column late phase.  The cartoon image (j),(k) and (l) are adapted from \citet{Sterling_Moore_2004}.\label{fig:figure2}}
\end{figure}

We turn to imaging observation such as SDO/AIA and STEREO/EUVI.
AIA instrument provides the corresponding imaging observations at 13.1 nm, 9.4 nm and 33.5 nm.
After checking the source region, we select out a region restricted to a small
region (figure~\ref{fig:figure2}(h)) where emission is mainly caused by the main flaring  loops and a larger region 
 mainly responsible for the emission of the late phase. The regions are depicted as the boxes in figure~\ref{fig:figure2}(h), over which the intensity profiles at different channels are plotted in figure~\ref{fig:figure1}(b). Obviously, the light curves from the main phase region (solid lines) behave in a similar way - they rise up, reach the peak, and recovery to the background nearly synchronously. The very close peak times of 13.1 nm (10 MK) to 33.5 nm (2.5 MK) signifies a rapid cooling process of the main phase.  In contrast,
the brightness of the late phase region (dashed lines in Figure~\ref{fig:figure1}(b)) shows a much more gradual evolution. The delay time  is 56 minutes for the AIA 13.1 nm, 128 minutes for AIA 9.4 nm, and 204 minutes for  AIA 33.5 nm (table ~\ref{tb:table1}). There also appears a dimming from 00:17 to 02:00 UT at the 33.5 nm (figure~\ref{fig:figure1}(b), blue dashed line) due to depletion of plasma density in areas associated with eruptions.  These observations coincide with previous conclusion that two set of loop systems exist in the active region. Different from the compact loop system for the flare main phase, the late phase
originates from another set of higher and longer loops.

\citet{Chen_etal_2016} analyzed the morphology of this case in detail and argued that the overall configuration of the AR is a quadrupolar magnetic configuration.
Figure~\ref{fig:figure2} shows the evolution of the eruption in different view angles during different stages with pseudo 3D schematic at the bottom. 
Here all the images have been rotated with the eruption direction towards to the top in order to compare with the cartoon. 
STEREO\_A EUVI 19.5 nm provides the disk observation (Figure~\ref{fig:figure2}(a)-(c)), STEREO\_B 19.5 nm (Figure~\ref{fig:figure2}(d)-(f)) and SDO 9.4 nm
(Figure~\ref{fig:figure2}(g)-(i)) provide limb views. Two sidelobe structures can be easily identified at the AIA 9.4 nm channel, indicating that hot structures  similar to the bipolar configuration on both sides in figure~\ref{fig:figure2}(j) cartoon. During the flare main phase,  a core structure expands outward, pushes the sidelobes aside and finally erupts into the interplanetary space as seen in hot channels (like AIA 9.4 nm in figure~\ref{fig:figure2}(g,h), figure~\ref{fig:figure3} and the accompanied movie). The cartoon in figure~\ref{fig:figure2}(k) depicts this scenario, but the low temperature coverage of EUVI 19.5 nm channel makes it only see the brightness of the post flare loops (PFLs), with their footpoints marked as P1 and P2. 
When it comes to the late phase stage, a totally different loop system appears, much longer and higher loops form with remote footpoints labeled as P3,P4 (figure~\ref{fig:figure2}(c) and (f)). Figure~\ref{fig:figure2}(i) shows the hot part of these loops on the top region. From figure~\ref{fig:figure1}(b) we could know that, the emission of the late phase mainly comes from the big box region, namely the late phase loop system. 
The late phase loop system is often considered as a pre-existing structure somehow connected magnetically with the main phase loop \citep{Liuk_etal_2013,Sun_etal_2013},  
but our observation seems to be in favor of another scenario - the late phase loop system comes from the newly reconnected large-scale arcades as the overlying  field above the quadrupolar configuration is stretched by the erupting CME and then reconnects as shown by Figure~\ref{fig:figure2}(i) \citep{Hock_etal_2012}.

\subsection{Late phase loop evolution} \label{subsec:evo}
\begin{figure}[ht!]
\plotone{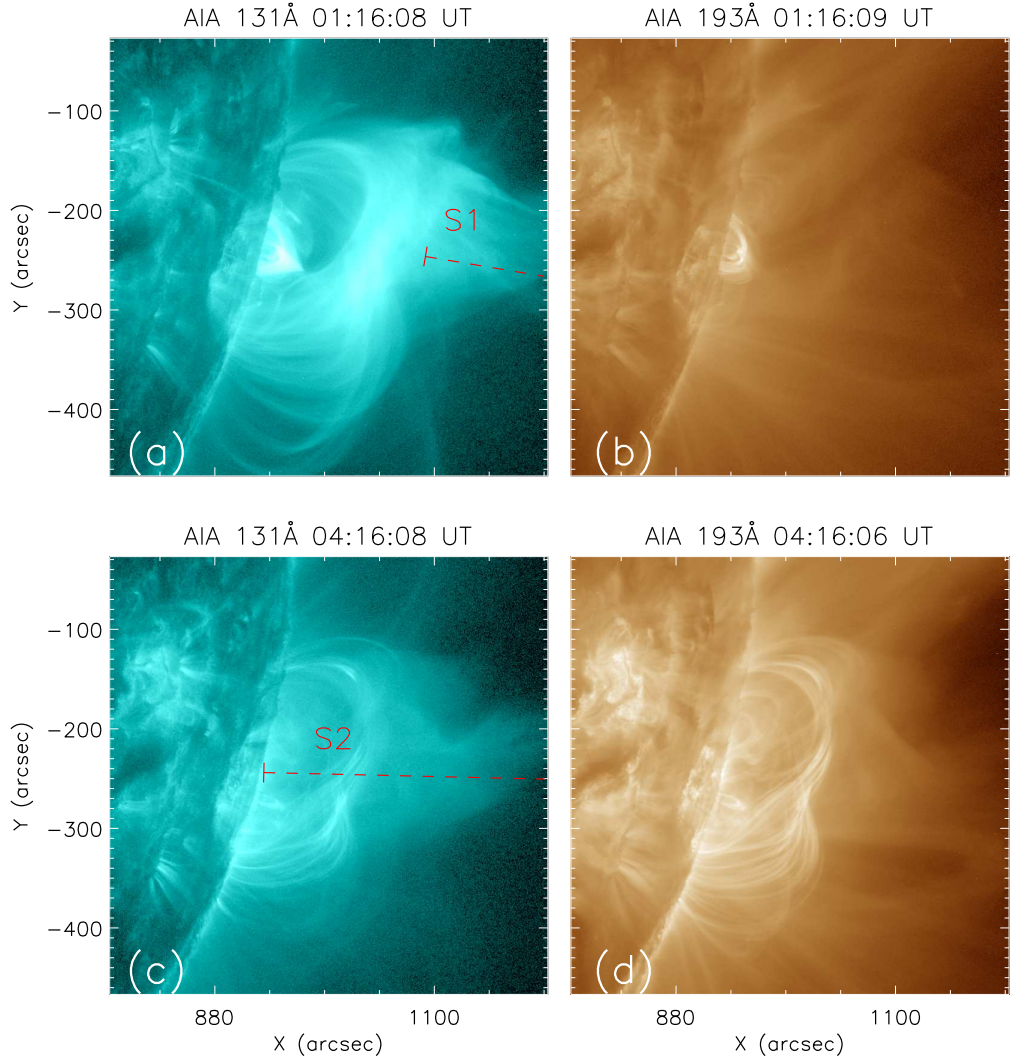}
\caption{The morphology and thermodynamics evolution during the late phase. Panel (a) shows the cusp structure on the top of late phase loops in AIA 13.1 nm, but in panel (b), only the post flare loops from the main phase region could be seen in AIA 19.3 nm; 
Panels (c) and (d)  shows that when these late phase loops cool down, in AIA 13.1 nm and 19.3 nm we can see large arcade loops overlying on the top. Here we consider that the observations in AIA 13.1 nm in panel (c)  mainly come from the Fe VIII line emission (0.4 MK) which is blended in AIA 13.1 nm. 
(An animation of this figure is available.)
\label{fig:figure3}}
\end{figure}

This event presents a novel perspective to the origin of the EUV late phase.  Interestingly, 
 unusual large and hot overlying arcades appear above the flare after the CME eruption as seen from the AIA view (figure~\ref{fig:figure3}(a)). This structure is invisible in the low temperature wavelengths like AIA 19.3 nm and EUVI\_B 19.5 nm.   Cusp-shaped loops appeared at the top of the arcades that rapidly retreat sunward, and continuously pile up above the arcade (Figure~\ref{fig:figure3}(a) and the accompanying animation). As these overlying arcades cool down,  numbers of loops sequentially show up from lower part to higher part (figure~\ref{fig:figure3}(d)) at the AIA 19.3 nm passband.  Due to a strong cold line (Fe VIII, 0.4 MK) blending, the AIA 13.1 nm passband also shows cold loops during the cooling process (figure~\ref{fig:figure3}(c)). Based on the morphology analysis in multiple-temperature, the evolution of the late phase loop system experienced three stages: initial stage, characterised by relatively low temperature high-lying arcades over the flare region (figure~\ref{fig:figure2}(j)); eruptive stage,  similar to the  classical flare magnetic reconnection scenario, in which the erupting CME stretches the overlying field, whose legs seem to curve-in toward the reconnection site; recovery stage, seen as the shrinkage of cusp structure in the high temperature channel (figure~\ref{fig:figure3}(a)), which is commonly considered as X-type configuration, as sketched by figure~\ref{fig:figure2}(i). 

\begin{figure}[ht!]
\plotone{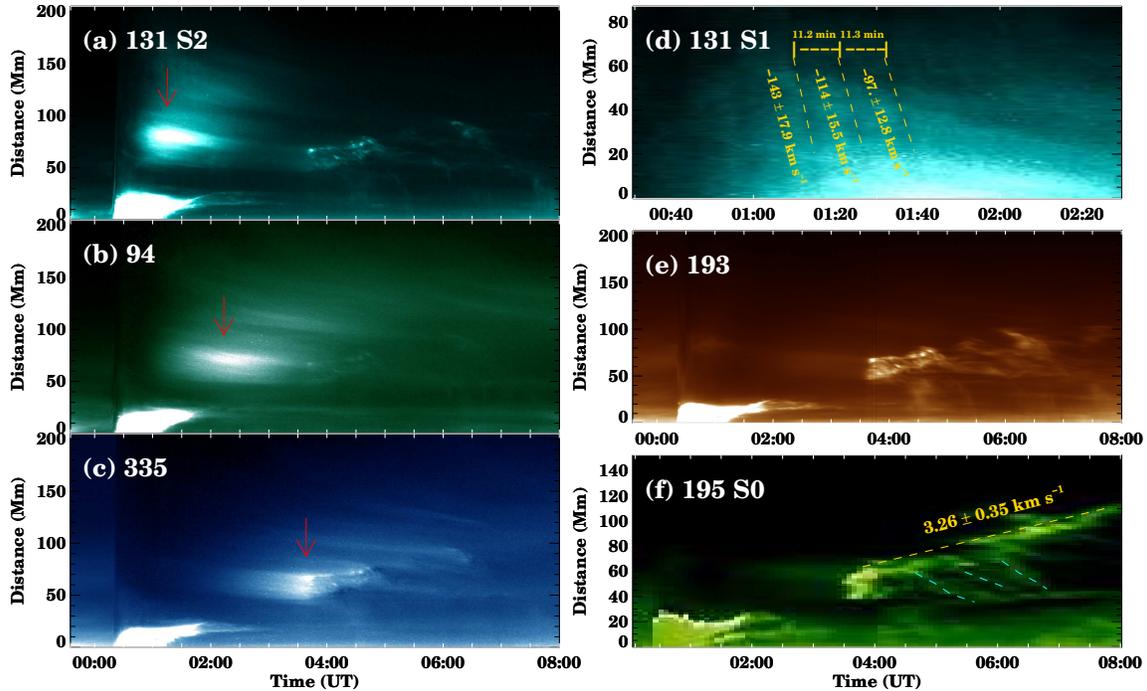}
\caption{Slice-Time stacking plots for slits S0-S2. (a)-(c) and (e) S2 plots at AIA 13.1 nm, 9.4 nm, 33.5 nm and 19.3 nm capture the evolution of the late phase loops. (d) S1 plot captures the center of the cusp structure. (f) S0 plot at EUVI\_B 19.5 nm provides another view angle to see the evolution of the late phase loops. The vertical arrows point out the brightness monments and the horizontal arrows mark out the  duration of brightness in each plot. \label{fig:figure4}}
\end{figure}

To further investigate the evolution of the late phase loop system, we perform a quantitative analysis. Here we study two evolution: shrinkage motion and cooling process of the late phase loops. We draw three slits: Slit S1 in figure~\ref{fig:figure3}(a) is located at the cusp location, while S0 in figure~\ref{fig:figure2}(f) and S2 in figure~\ref{fig:figure3}(c) indicate the eruption direction are used to trace late phase loops. Figure~\ref{fig:figure4}(a),(b) and (c) are the slice-time stacking plots of slit S2 at the AIA 13.1 nm, 9.4 nm and 33.5 nm passbands. A bright core region appears in all these channels with its height above 50 Mm.  The time delay of the brightness from 13.1 nm to 33.5 nm indicates the cooling of the overlying arcades. The vertical arrow in each channel gives the timing of the maximum intensity.  When the temperature reaches a low level (such as 1 MK ), in the slice-time diagram of slit S2 from AIA 19.3 nm channel and slit S0 from EUVI 19.5 nm channel, a bright feature  slowly rises up with a fitting speed of 3.26 km s$^{-1}$. For PFLs, it is generally recognized that this kind of rising motion is an apparent ascending of these overlying loops instead of real mass motion.  Here for the late phase loops, we think that the same process could explain this motion.
From figure~\ref{fig:figure4} and accompanied movie one can see that the late phase loops behave in the similar manner like PFLs, i.e., the newly reconnected loops continuously accumulate above the flare loops. So this gives us a hint that the duration of the brightness in each channel has direct correlation with the numbers of the shrinkage loops. 

In figure~\ref{fig:figure4}(d),  a series of dark features continuously drop down towards to the flare loop top. 
These voids known as supra-arcade downflows
(SADs) were first observed by Yohkoh SXT and TRACE\citep[e.g.,][]{McKenzie_Hudson_1999,Gallagher_etal_2002} and 
SADs are closely related to the outflows produced by magnetic reconnection occurring high
in the corona.
Thus the dark features could be a consequence of newly reconnected loops.  We identify the obvious trajectories of three dark features to trace the shrinkage motion.  The velocities range from 97 km s$^{-1}$ to 143 km s$^{-1}$. 
The intervals between the neighbouring stripes are 11.2 minutes and 11.3 minutes in the AIA 13.1 nm slice-time stacking plot of Figure~\ref{fig:figure4}(d), respectively.  
These stripes manifest the trajectories of SADs,
which could be due to the SADs retract from the reconnection sites high in the corona and continuously cool down, giving rise to consecutive brightness in the different AIA channels.

To quantify the loop length of single late phase loop for the later simulation. We used the IDL routine scc\_measure.pro in SSW to trace out the path of the loop in 3D space \citep{Thompson_2009,Zhou_etal_2017}.  Because  STEREO EUVI\_A/B only have cool line, we  use  the EUVI 19.5 nm ($\sim$ 1.5 MK) and  AIA 19.3 nm ($\sim$ 1.6 MK) to reconstruct the loop.  Both images in the AIA 19.3 nm and EUVI 19.5 nm at 04:10 UT show multiple clear bright loops. We select out a loop and obtain its loop half length of 138 Mm. Its initial length should be larger because it has retracted
before it becomes visible.

\subsection{EBTEL modeling of the late phase emissions with additional heating}\label{sec:mod}
By far there are two major explanations for the EUV late phase: one is additional heating, and the other is long-lasting cooling.  A definite answer perhaps requires modeling of a multitude of loops based on observations
\citep[e.g.][]{Qiu_etal_2012,Liuw_etal_2013,Hock_etal_2012}. Here, from possible signatures of the reconnection in the high corona, we take a heuristic approach with additional heating to model the late phase.
The EBTEL-based flare model\citep{Hock_etal_2012} is employed to simulate radiative output of the late phase loop system. This model uses multiple EBTEL loops (e.g., 22 EBTEL loops) with variable loop lengths and heating rate profiles to synthesize the light curves of several EUV lines as measured by EVE, 13 out of 22 loops are employed to model the late phase emissions. Here we use 10 EBTEL loops to model the late phase radiative output.


\subsubsection{the EBTEL model}
 For a single EBTEL loop, its evolution can be controlled by loop-averaged equation of continuity and energy \citep{Klimchuk_etal_2008, Cargill_etal_2012}:
\begin{equation}
\frac{dn}{dt}=-\frac{c_2}{5c_3k_BT}[\frac{F_c}{L}+c_1n^2\Lambda(T)] \,,
\end{equation}
and

\begin{equation}
\frac{dP}{dt}=\frac{2}{3}[Q(t)-(1+c_1)n^2\Lambda(T)]\,,
\end{equation}

where $n$, $P$, and $T$ are the mean density, pressure, and temperature
of the loop, respectively, which follow from the equation of state, $P = 2nk_BT$ with $k_B$ being the Boltzmann
constant, $c_2$ (= 0.9) is the typical ratio of average
to apex temperature of a coronal loop, $c_3$ (= 0.6) is the ratio between coronal base temperature and apex temperature, $Q(t)$ is the
volumetric heating rate, $\Lambda(T)$ is the optically thin radiative loss function,
$F_c =-(2/7) \kappa_0 (T /c_2 )^{7/2}/l$ ($l$ the half loop length) is the thermal conductive flux,
and $c_1$ is the ratio of radiative loss rate of the TR to that of the corona.

In addition, we only consider thermal heating $Q(t)$ and ignore non-thermal electron beam heating. 
The rationality of predigestion mainly relys on following reasons.
In the additional heating scenario, the non-thermal heating is unlikely to be the major heating source during the late phase based on previous study that the late-phase arcades do not show obvious HXR emission\citep[e.g.][]{Li_etal_2012,Li_etal_2014,Sun_etal_2013}. In fact, \citet{Li_etal_2014} checked the effect of non-thermal beam heating using a flux of which the total energy comparable with that of thermal heating, and found that the non-thermal effect is not significant in the EUV late phase. 
Thermal heating primarily raises the temperature of the coronal plasma, where non-thermal beam heating primarily raises the density.   
\citet{Liuw_etal_2013} discussed that the increased density caused by the non-thermal beam could somewhat enhance the EUV late-phase emission, but it affects the timing of the EUV late phase much less than a loop length variation. Meanwhile we only calculate the emission from the coronal part, because the footpoints of these loops are located behind the solar limb.


The EBTEL code is publicly accessible at \url{https://github.com/rice-solar-physics/EBTEL}. The main program is called ebtel2.pro that computes the evolution of spatially-averaged loop quantities using the above simplified equations. Combined with the CHIANTI atomic database \citep{Dere_etal_1997,Dere_etal_2009}, the routine - intensity\_ebtel.pro - computes the line intensity as a function of time.

\subsubsection{EBTEL-based flare Model}
There are three steps to calculate the line emission based on the EBTEL model. First, for each single loop, EBTEL program gives the differential emission measure (DEM) as a function of time and temperature. The inputs of the model are heating function, $Q(t)$ (erg cm$^{-3}$ s$^{-1}$) and loop half-length (cm).  The heating function has a linear slope, i.e.
\begin{equation}
 Q(t) = 
  \begin{cases} 
   \ Q_{bkgd}+Q_0+\frac{2Q_0}{dt}(t-t_0) & \text{if } t_0-\frac{dt}{2} < t < t_0 \,,\\
   \ Q_{bkgd}+Q_0-\frac{2Q_0}{dt}(t-t_0) & \text{if } t_0 \leq t < t_0+\frac{dt}{2}\,,
  \end{cases}
\end{equation}

Where $Q_0$ is the heating amplitude, $dt$ is the heating duration, which is set to be the interval between two time-adjacent loops, namely 11.2 minutes, and $Q_{bkgd}$ is the background heating at a low level of $10^{-6}$ erg cm$^{-3}$ s$^{-1}$.  The initial setting is similar to that in \citet{Hock_etal_2012} except that here we fix the start time of the heating ($t_0$) for each loop based on observations:
\begin{equation}
t_0=t\sp{\prime}_0+i \times dt\,,
\end{equation}
Where $t\sp{\prime}_0$ is the first episode of the late phase heating.  As the reconnection continues, higher reconnected loops will shrink backward, and the half-length of these loops, $l$, will increases.  Here we simply  assume a linear increase in length of consecutive loops and a linear decrease in the heating magnitude.

EBTEL will return $DEM\_COR(t,T)$ (DEM from the corona part) and $DEM\_TR(t,T)$ (DEM from the transition region). Due to occulting effect of the solar limb, here we only need to calculate the emission from the corona.  The line radiance is the convolution of the DEM with the contribution function $G(T)$  that contains abundance factor, i.e.
\begin{equation}
I_{loop}(t)=\int_T DEM(t,T)G(T)dT\,.
\label{eqn:Iloop}
\end{equation}


Note that the contribution function is calculated by the program gofnt.pro in CHIANTI package. 

At last, the total emission of the late phase is generated by combining the contribution of all EBTEL loops. To compare with EVL data,  a constant normalization factor C is used to convert the  radiation at Sun to the irradiance received at Earth (1AU) assuming a uniform (constant) angular distribution of flare energy release, as showed below:

\begin{equation}
E_{model}(t)=C \sum_{i=1}^{N_{loops}}A_iI_{loop,i}(t)\,,
\label{eqn:Emis}
\end{equation}
and
\begin{equation}
C=(0.001\frac{W/m^2}{ergs/cm^2/s})d{\Omega}\,,
\label{eqn:factor}
\end{equation}
Where $A_i$ is loop cross-section area, which is the same for the modeling loops.  $N_{loops}$ is the number of the loops, and $d\Omega$ ($3.56 \times 10^{-24}$ sr m$^{-2}$) is the solid angle subtended by 1 m$^2$ at 1 AU. Which yields a normalization factor C  of $3.56 \times 10^{-27}$\citep{Hock_2012}.
All parameters in the EBTEL-based flare model are listed in table ~\ref{tb:table2}.
 
\begin{table}[]
\centering
\caption{Late Phase Parameters for the EBTEL-based Flare Model}
\label{tb:table2}
\begin{tabular}{@{}|lllll|@{}}
\toprule
                                        & Parameters     & Our model                             & Hock model                         & Note \\ \midrule
\multirow{5}{*}{Single loop parameters} & $l^*$          & $l\sp{\prime}_0+i \times \Delta l$    & free                               & Loop half-length (cm)   \\
                                        & $t0$           & $t\sp{\prime}_0+i \times \Delta t$    & free                               & Time of peak impulsive heating (s),$\Delta t = 672$ s   \\
                                        & $dt$           & 672s                                  & 300s                               & Duration of heating (s)  \\
                                        & $Q0$           & $Q\sp{\prime}_0+i \times \Delta h$    & free                               & Amplitude of heating (erg\ cm$^{-3}$ \ s$^{-1}$) \\
                                        & $A$            & free                                  & free                               & Loop cross-section area (cm$^2$)
     \\ \midrule
\multirow{2}{*}{Global parameters}      & $Nloops$       & 10                                    & 22                                 & Number of EBTEL loops    \\
                                        & $C$            &$3.56 \times 10^{-27}$                 & $3.56 \times 10^{-27}$             & Normalization factor (sr \ m$^{-2}$)  \\ \bottomrule
\end{tabular}
\raggedright

The loop half-length ($l^*$) is free parameter that is constrained using the imaging observations,  $l\sp{\prime}_0$, $t\sp{\prime}_0$ and $Q\sp{\prime}_0$ are free paramters.  Numbers of loops for our model is estimated from observation. 
\end{table}

\subsubsection{Modeling Result}
\begin{figure}[ht!]
\plotone{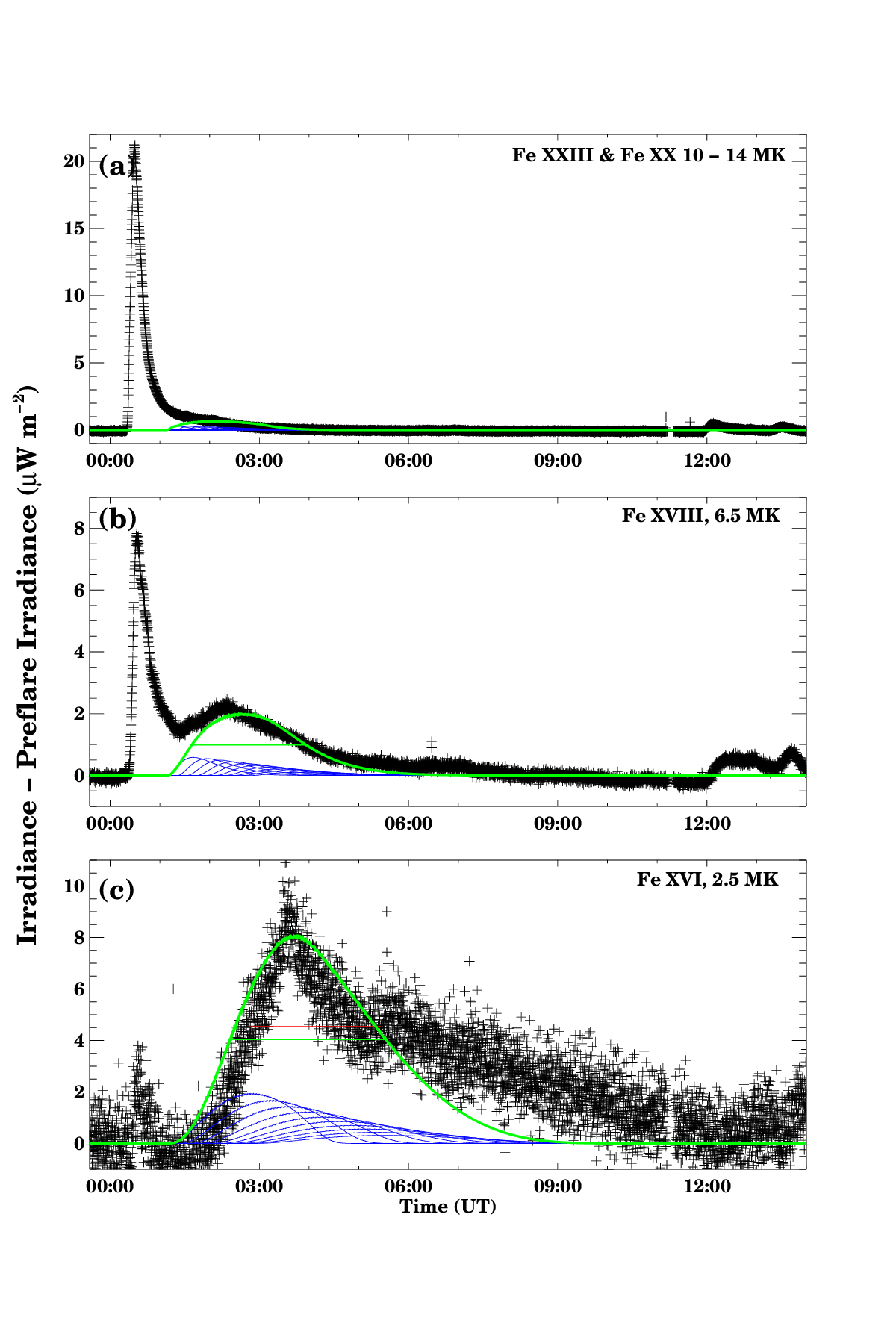}
\caption{Results of the EBTEL-based flare model for both
the Fe XVIII (middle) and the Fe XVI line (bottom). The observation and modeling of the Fe XX/XXIII blended line are placed on the top just for  
reference. 
The pluses are the EVE observations with the pre-flare background
irradiance subtracted off. The
blue lines are the contribution from each individual coronal loop
strand, the green line is the output from
the model. 
\label{fig:figure5}}
\end{figure}

Using the EBTEL-based flare Model, we reproduce the irradiance in different emission lines for the late phase loop system and compare it with EVE's observations (figure~\ref{fig:figure5}).
The late phase emission in hot line (13.3 nm, $>10$ MK)  could merely be seen as  the late phase emission is largely submerged under main phase emission. This blending effect also affects the 9.4 nm EVE line, resulting in an obvious overlapping of the main phase and late phase emissions during 01:00 UT to 02:00 UT (as shown by the black line in figure~\ref{fig:figure5}(b)).  So we only use the 9.4 nm data after 02:00 UT to do the fitting. As for 33.5 nm line irradiance, the late -phase bump is clearly identified.

We fit both 9.4 nm and 33.5 nm lines using MPFIT package. The result shows a good agreement between the observations and modeling (figure~\ref{fig:figure5}). The observed 13.3 nm emission and modeled one are plotted in the top panel of figure~\ref{fig:figure5}. The best-fit gives an initial loop half-length ($l\sp{\prime}_0$) of 179 Mm and heating rate ($Q\sp{\prime}_0$) of $0.022$ erg cm$^{-3}$ s$^{-1}$, of which comparable to that in \citet{Hock_etal_2012}'s model (
$\sim 0.01$ erg cm$^{-3}$ s$^{-1}$), 
but less than that ($0.2$ erg cm$^{-3}$ s$^{-1}$) of case 1 in \citet{Li_etal_2012} with one order of magnitude. However, in \citet{Li_etal_2012}'s model, only one 
loop was used to model the late phase loop.
the best-fit loop cross-section is $1.3 \times 10^{16}$ cm$^2$ corresponding to a loop width of 1.3 Mm.

\begin{table}[!htbp] 
\centering 
 \caption{Comparison of Properties from Observations and Modelings}\label{tb:table3}
\begin{tabular}{@{\extracolsep{1pt}} cccccccccc} 

\toprule

{\centering Passband}              & Ion & Temperature  & \multicolumn{3}{c}{Observation}         && \multicolumn{3}{c}{modeling} \\ 
\cmidrule{4-6} \cmidrule{8-10}
                                   &     &              & Peak Intensity& Peak Time&FWHM          &&  Peak Intensity &Peak Time&FWHM  \\
                               (nm)&     &(MK)          & $(\mu$W)     & (UT)     &(minutes)     &&  $(\mu$W)      & (UT)     &(minutes)  \\

\midrule

9.4              & Fe XVIII    &6.5        & 2.000    & 02:26:02 & *        &   & 1.985    & 02:39:02 & 140  \\ 
33.5             & Fe XVI      &2.5        & 9.071    & 03:31:42 & 158      &   & 8.071    & 03:42:32 & 192  \\ 

\hline \\[-1.8ex] 
\end{tabular}\\
\raggedright

\end{table} 
Using the EBTEL model, we reconstruct the irradiance of the flare late phase and study the role of additional heating in the appearance of an EUV late phase. Our results show that the late phase peak times in observations and modeling for each line  are very close with each other (within 15 minutes). Hence the long delay appearing in the warm lines can be well explained. 
\citet{Liuk_etal_2013} estimated the cooling time of the late phase loop arcade by using the formula derived by \citet{Cargill_etal_1995} based on a simple cooling model:
\begin{equation}
\tau_{cool} \doteq 2.35 \times 10^{-2} l^{5/6} T_e^{-1/6} n_e^{-1/6}\,.
\label{eqn:cool}
\end{equation}
The parameters $l$, $T_e$, and $n_e$ are, respectively, the loop half-length, electron temperature, and electron density at the beginning of the cooling.
Here we use the best-fit loop half-length (179 Mm) as the input. 
The temperature and density are determined using differential emission measure (DEM) analysis\citep{Schmelz_etal_2011,Cheng_etal_2012}. 
The derived initial temperature of the late phase is 14 MK and density is $\sim 6 \times 10^9$ cm$^{-3}$. These parameters yield a cooling time around 210 minutes. However, the delay time of the late phase peak (at 05:00 UT) in Fe IX 17.1 nm (0.7 MK) is about 280 minutes.
The late appearance of the late phase loop arcade in the EUV band is mainly a cooling-delay effect with the additional consecutive  heating. 



\section{Summary and Discussion}\label{sec:dis}

In summary, we examine an extremely large late phase during an eruptive solar flare.
The magnetic topology is a quadrupolar magnetic configuration, which facilitates the coexistence
of lower core field and high large-scale overlying field. The cusp structure observed in the hot channel evidences the occurrence of magnetic reconnection process\citep{Tsuneta_etal_1992,Aschwanden_2005},  the downward motion of the hot loops  represents the shrinkage of newly formed loops\citep{Forbes_Acton_1996}. All these observations suggest the  possibility of magnetic reconnection taking place high above. Such a reconnection process is caused by the erupting CME. The overlying arcades are stretched upward by the CME, and the stretched field lines re-close after the reconnection beneath the CME. This process  ,similar to the tether-cutting reconnection, happens in the late phase stage. It provides additional heating mechanism for the EUV late phase.  The detailed evolution for the late phase loop system is described as follows: 

\begin{enumerate}
\item Before the flare, the large-scale overlying arcades connects the outer positive and negative polarities. The quadrupolar magnetic configuration distinguishes the overlying arcades from lower core field.
\item When the CME erupts, the overlying arcades are stretched outward, the tether-cutting-like reconnection takes place between two legs of the overlying field.   The reconnection plays two roles: Magnetic topology is rearranged and magnetic energy is converted to additional heating that is transferred into the late-phase loop system. 
\item As the reconnection continues, the cusp structure retreats backward and show some quasi-periodic characteristics. 
\item As time goes on,  these loops cool down and become visible in the low temperature lines, such as the AIA 19.3 nm and EUVI 19.5 nm. Because  inner loops form first and then cool down first,  like the post-flare loop systems, the loop top shows up a rising motion.
\end{enumerate}

As a summary, we consider the late-phase emission comes from a set of gradually downward moving loops rather than a pre-existing loop system.  The observational  features also imply the existence of additional heating of the large-scale flare loops, which is shown in the bottom pannel of figure~\ref{fig:figure2}.  

We also use EBTEL to model the EVE irradiance in several lines. The modeling result shows well agreement with the observations.  The peak time in each line is very close to the observations, indicating that additional heating gives rise to the delayed appearance of the late phase.

\citet{Liu_etal_2015} and \citet{Wang_etal_2016} argued that the extremely large late phase emission as the energy re-deposition from the failed flux rope into the thermal emissions, where \citet{Dai_etal_2018b} presented a more complicated scenario in their case: The late-phase loops are mainly produced by the first-stage QSL reconnection, while the second-stage reconnection is
responsible for the heating of main flaring loops. However, the production of extremely large EUV late phase in this flare is from the late phase reconnection occurring high above the main flare region. The CME originated from main flare is only the trigger.

We acknowledge the use of the data from GOES, from HMI and AIA instruments onboard
SDO, and from EUVI instruments onboard STEREO. 
Z.J.Z. is supported by the grants from the Open Project of CAS Key Laboratory of Geospace Environment.
X.C. are supported by NSFC under grants 11722325, 11733003, 11790303, and by Jiangsu NSF under grant BK20170011.
D.Y. is supported by National Natural Science Foundation of China under grants 11533005, and 973 Project of China under grant 2014CB744203. 
Y.M.W. is supported by the grants from NSFC (41574165 and 41774178).  
J.C. is supported by NSFC through grants 41525015,41774186. 

\bibliographystyle{agu}
\bibliography{citation}

\begin{thebibliography}{}
\expandafter\ifx\csname natexlab\endcsname\relax\def\natexlab#1{#1}\fi

\bibitem[{{Antiochos} \& {Sturrock}(1978)}]{Antiochos_Sturrock_1978}
{Antiochos}, S.~K., \& {Sturrock}, P.~A. 1978, \apj, 220, 1137

\bibitem[{{Aschwanden}(2005)}]{Aschwanden_2005}
{Aschwanden}, M.~J. 2005, {Physics of the Solar Corona. An Introduction with
  Problems and Solutions (2nd edition)}

\bibitem[{{Brueckner} {et~al.}(1995){Brueckner}, {Howard}, {Koomen},
  {Korendyke}, {Michels}, {Moses}, {Socker}, {Dere}, {Lamy}, {Llebaria},
  {Bout}, {Schwenn}, {Simnett}, {Bedford}, \& {Eyles}}]{Brueckner_etal_1995}
{Brueckner}, G.~E., {Howard}, R.~A., {Koomen}, M.~J., {et~al.} 1995, \solphys,
  162, 357

\bibitem[{Bumba \& Krivsk{\'{y}}(1959)}]{Bumba_Krivsky_1959}
Bumba, V., \& Krivsk{\'{y}}, L. 1959, Bulletin of the Astronomical Institutes
  of Czechoslovakia, 10, 221

\bibitem[{{Cargill} {et~al.}(2012){Cargill}, {Bradshaw}, \&
  {Klimchuk}}]{Cargill_etal_2012}
{Cargill}, P.~J., {Bradshaw}, S.~J., \& {Klimchuk}, J.~A. 2012, \apj, 752, 161

\bibitem[{{Cargill} {et~al.}(1995){Cargill}, {Mariska}, \&
  {Antiochos}}]{Cargill_etal_1995}
{Cargill}, P.~J., {Mariska}, J.~T., \& {Antiochos}, S.~K. 1995, \apj, 439, 1034

\bibitem[{Chen {et~al.}(2016)Chen, Du, Zhao, Wu, Liu, Wang, Ruan, Feng, \&
  Song}]{Chen_etal_2016}
Chen, Y., Du, G., Zhao, D., {et~al.} 2016, The Astrophysical Journal Letters,
  820, L37

\bibitem[{Cheng {et~al.}(1985)Cheng, Pallavicini, Acton, \&
  Tandberg-Hanssen}]{Cheng_etal_1985}
Cheng, C.~C., Pallavicini, R., Acton, L.~W., \& Tandberg-Hanssen, E. 1985, The
  Astrophysical Journal, 298, 887

\bibitem[{{Cheng} {et~al.}(2012){Cheng}, {Zhang}, {Saar}, \&
  {Ding}}]{Cheng_etal_2012}
{Cheng}, X., {Zhang}, J., {Saar}, S.~H., \& {Ding}, M.~D. 2012, \apj, 761, 62

\bibitem[{Contarino {et~al.}(2003)Contarino, Romano, Yurchyshyn, \&
  Zuccarello}]{Contarino_etal_2003}
Contarino, L., Romano, P., Yurchyshyn, V.~B., \& Zuccarello, F. 2003, Solar
  Physics, 216, 173

\bibitem[{Dai \& Ding(2018)}]{Dai_etal_2018a}
Dai, Y., \& Ding, M. 2018, The Astrophysical Journal, 857, 99

\bibitem[{{Dai} {et~al.}(2018){Dai}, {Ding}, {Zong}, \&
  {Yang}}]{Dai_etal_2018b}
{Dai}, Y., {Ding}, M., {Zong}, W., \& {Yang}, K.~E. 2018, \apj, 863, 124

\bibitem[{Dai {et~al.}(2013)Dai, Ding, \& Guo}]{Dai_etal_2013}
Dai, Y., Ding, M.~D., \& Guo, Y. 2013, The Astrophysical Journal Letters, 773,
  L21

\bibitem[{{Dere} {et~al.}(1997){Dere}, {Landi}, {Mason}, {Monsignori Fossi}, \&
  {Young}}]{Dere_etal_1997}
{Dere}, K.~P., {Landi}, E., {Mason}, H.~E., {Monsignori Fossi}, B.~C., \&
  {Young}, P.~R. 1997, \aaps, 125, 149

\bibitem[{{Dere} {et~al.}(2009){Dere}, {Landi}, {Young}, {Del Zanna},
  {Landini}, \& {Mason}}]{Dere_etal_2009}
{Dere}, K.~P., {Landi}, E., {Young}, P.~R., {et~al.} 2009, \aap, 498, 915

\bibitem[{F{\'{a}}rn{\'{i}}k {et~al.}(2003)F{\'{a}}rn{\'{i}}k, Hudson,
  Karlick{\'{y}}, \& Kosugi}]{Farnik_etal_2003}
F{\'{a}}rn{\'{i}}k, F., Hudson, H., Karlick{\'{y}}, M., \& Kosugi, T. 2003,
  $\backslash$Aap, 399, 1159

\bibitem[{{Forbes} \& {Acton}(1996)}]{Forbes_Acton_1996}
{Forbes}, T.~G., \& {Acton}, L.~W. 1996, \apj, 459, 330

\bibitem[{{Gallagher} {et~al.}(2002){Gallagher}, {Dennis}, {Krucker},
  {Schwartz}, \& {Tolbert}}]{Gallagher_etal_2002}
{Gallagher}, P.~T., {Dennis}, B.~R., {Krucker}, S., {Schwartz}, R.~A., \&
  {Tolbert}, A.~K. 2002, \solphys, 210, 341

\bibitem[{{Hock}(2012)}]{Hock_2012}
{Hock}, R.~A. 2012, PhD thesis, University of Colorado at Boulder

\bibitem[{{Hock} {et~al.}(2012){Hock}, {Woods}, {Klimchuk}, {Eparvier}, \&
  {Jones}}]{Hock_etal_2012}
{Hock}, R.~A., {Woods}, T.~N., {Klimchuk}, J.~A., {Eparvier}, F.~G., \&
  {Jones}, A.~R. 2012, ArXiv e-prints, arXiv:1202.4819

\bibitem[{{Hudson}(2011)}]{Hudson_2011}
{Hudson}, H.~S. 2011, \ssr, 158, 5

\bibitem[{{Kaiser} {et~al.}(2008){Kaiser}, {Kucera}, {Davila}, {St.~Cyr},
  {Guhathakurta}, \& {Christian}}]{Kaiser_etal_2008}
{Kaiser}, M.~L., {Kucera}, T.~A., {Davila}, J.~M., {et~al.} 2008, \ssr, 136, 5

\bibitem[{{Klimchuk} {et~al.}(2008){Klimchuk}, {Patsourakos}, \&
  {Cargill}}]{Klimchuk_etal_2008}
{Klimchuk}, J.~A., {Patsourakos}, S., \& {Cargill}, P.~J. 2008, \apj, 682, 1351

\bibitem[{Lemen {et~al.}(2012)Lemen, Title, Akin, Boerner, Chou, Drake, Duncan,
  Edwards, Friedlaender, Heyman, Hurlburt, Katz, Kushner, Levay, Lindgren,
  Mathur, McFeaters, Mitchell, Rehse, Schrijver, Springer, Stern, Tarbell,
  Wuelser, Wolfson, Yanari, Bookbinder, Cheimets, Caldwell, Deluca, Gates,
  Golub, Park, Podgorski, Bush, Scherrer, Gummin, Smith, Auker, Jerram, Pool,
  Soufli, Windt, Beardsley, Clapp, Lang, \& Waltham}]{Lemen_etal_2012}
Lemen, J.~R., Title, A.~M., Akin, D.~J., {et~al.} 2012, Solar Physics, 275, 17

\bibitem[{Li {et~al.}(2014)Li, Ding, Guo, \& Dai}]{Li_etal_2014}
Li, Y., Ding, M.~D., Guo, Y., \& Dai, Y. 2014, The Astrophysical Journal, 793,
  85

\bibitem[{{Li} {et~al.}(2012){Li}, {Qiu}, \& {Ding}}]{Li_etal_2012}
{Li}, Y., {Qiu}, J., \& {Ding}, M.~D. 2012, \apj, 758, 52

\bibitem[{Liu {et~al.}(2015)Liu, Wang, Zhang, Cheng, Liu, \&
  Shen}]{Liu_etal_2015}
Liu, K., Wang, Y., Zhang, J., {et~al.} 2015, The Astrophysical Journal, 802, 35

\bibitem[{Liu {et~al.}(2013a)Liu, Zhang, Wang, \& Cheng}]{Liuk_etal_2013}
Liu, K., Zhang, J., Wang, Y., \& Cheng, X. 2013a, The Astrophysical Journal,
  768, 150

\bibitem[{{Liu} {et~al.}(2013b){Liu}, {Qiu}, {Longcope}, \&
  {Caspi}}]{Liuw_etal_2013}
{Liu}, W.-J., {Qiu}, J., {Longcope}, D.~W., \& {Caspi}, A. 2013b, \apj, 770,
  111

\bibitem[{Martin(1980)}]{Martin_1980}
Martin, S.~F. 1980, Solar Physics, 68, 217

\bibitem[{{Masson} {et~al.}(2017){Masson}, {Pariat}, {Valori}, {Deng}, {Liu},
  {Wang}, \& {Reid}}]{Masson_etal_2017}
{Masson}, S., {Pariat}, {\'E}., {Valori}, G., {et~al.} 2017, \aap, 604, A76

\bibitem[{{McKenzie} \& {Hudson}(1999)}]{McKenzie_Hudson_1999}
{McKenzie}, D.~E., \& {Hudson}, H.~S. 1999, \apjl, 519, L93

\bibitem[{Pesnell {et~al.}(2012)Pesnell, Thompson, \&
  Chamberlin}]{Pesnell_etal_2012}
Pesnell, W., Thompson, B., \& Chamberlin, P. 2012, $\backslash$solphys, 275, 3

\bibitem[{{Qiu} {et~al.}(2012){Qiu}, {Liu}, \& {Longcope}}]{Qiu_etal_2012}
{Qiu}, J., {Liu}, W.-J., \& {Longcope}, D.~W. 2012, \apj, 752, 124

\bibitem[{{Schmelz} {et~al.}(2011){Schmelz}, {Worley}, {Anderson}, {Pathak},
  {Kimble}, {Jenkins}, \& {Saar}}]{Schmelz_etal_2011}
{Schmelz}, J.~T., {Worley}, B.~T., {Anderson}, D.~J., {et~al.} 2011, \apj, 739,
  33

\bibitem[{{Stenborg} {et~al.}(2008){Stenborg}, {Vourlidas}, \&
  {Howard}}]{Stenborg_etal_2008}
{Stenborg}, G., {Vourlidas}, A., \& {Howard}, R.~A. 2008, \apj, 674, 1201

\bibitem[{{Sterling} \& {Moore}(2004)}]{Sterling_Moore_2004}
{Sterling}, A.~C., \& {Moore}, R.~L. 2004, \apj, 613, 1221

\bibitem[{Sun {et~al.}(2013)Sun, Hoeksema, Liu, Aulanier, Su, Hannah, \&
  Hock}]{Sun_etal_2013}
Sun, X., Hoeksema, J.~T., Liu, Y., {et~al.} 2013, The Astrophysical Journal,
  778, 139

\bibitem[{{Thompson}(2009)}]{Thompson_2009}
{Thompson}, W.~T. 2009, \icarus, 200, 351

\bibitem[{{Tsuneta} {et~al.}(1992){Tsuneta}, {Hara}, {Shimizu}, {Acton},
  {Strong}, {Hudson}, \& {Ogawara}}]{Tsuneta_etal_1992}
{Tsuneta}, S., {Hara}, H., {Shimizu}, T., {et~al.} 1992, \pasj, 44, L63

\bibitem[{{Van Hoven} \& Hurford(1984)}]{Van_Hurford_1984}
{Van Hoven}, G., \& Hurford, G.~J. 1984, Advances in Space Research, 4, 95

\bibitem[{{Wang} {et~al.}(2016){Wang}, {Zhou}, {Zhang}, {Liu}, {Liu}, {Shen},
  \& {Chamberlin}}]{Wang_etal_2016}
{Wang}, Y., {Zhou}, Z., {Zhang}, J., {et~al.} 2016, \apjs, 223, 4

\bibitem[{Warren \& Warshall(2001)}]{Warren_Warshall_2001}
Warren, H.~P., \& Warshall, A.~D. 2001, The Astrophysical Journal, 560, L87

\bibitem[{{Woods} {et~al.}(2006){Woods}, {Kopp}, \&
  {Chamberlin}}]{Woods_etal_2006}
{Woods}, T.~N., {Kopp}, G., \& {Chamberlin}, P.~C. 2006, Journal of Geophysical
  Research (Space Physics), 111, A10S14

\bibitem[{{Woods} {et~al.}(2011){Woods}, {Hock}, {Eparvier}, {Jones},
  {Chamberlin}, {Klimchuk}, {Didkovsky}, {Judge}, {Mariska}, {Warren},
  {Schrijver}, {Webb}, {Bailey}, \& {Tobiska}}]{woods_etal_2011}
{Woods}, T.~N., {Hock}, R., {Eparvier}, F., {et~al.} 2011, \apj, 739, 59

\bibitem[{Woods {et~al.}(2012)Woods, Eparvier, Hock, Jones, Woodraska, Judge,
  Didkovsky, Lean, Mariska, Warren, McMullin, Chamberlin, Berthiaume, Bailey,
  Fuller-Rowell, Sojka, Tobiska, \& Viereck}]{Woods_etal_2012}
Woods, T.~N., Eparvier, F.~G., Hock, R., {et~al.} 2012, Solar Physics, 275, 115

\bibitem[{{Zhou} {et~al.}(2017){Zhou}, {Zhang}, {Wang}, {Liu}, \&
  {Chintzoglou}}]{Zhou_etal_2017}
{Zhou}, Z., {Zhang}, J., {Wang}, Y., {Liu}, R., \& {Chintzoglou}, G. 2017,
  \apj, 851, 133

\end{thebibliography}

\clearpage



 \end{CJK*}
\end{document}